\documentclass[twocolumn,prl,color,superscriptaddress,psfig]{revtex4}

\def\beq{\begin{equation}}
\def\eeq{\end{equation}}
\def\123{YBa$_2$Cu$_3$O$_{6+x}$}
\def\214{La$_{2-x}$Sr$_x$CuO$_{4}$}
\usepackage{graphicx}
\usepackage{epsfig}
\usepackage{color,amsmath,graphicx,latexsym,psfig}

\begin{document} 

\title{NMR and LDA evidence for  spiral magnetic order in the
  chain cuprate LiCu$_2$O$_2$ }
\author{A.A. Gippius}
\affiliation{Moscow State University, 119899, Moscow, Russia}
\author{E.N. Morozova}
\affiliation{Moscow State University, 119899, Moscow, Russia}
\author{A.S. Moskvin}
\affiliation{Ural State University, 620083, Ekaterinburg, Russia}
\author{A.V. Zalessky}
\affiliation{Institute of Crystallography, RAS, 117333, Moscow, Russia}
\author{A.A. Bush}
\affiliation{Moscow Institute of Radiotechnics,  Electronics 
and Automation, 117464, Moscow, Russia}
\author{M. Baenitz}
\affiliation{Max Planck Institut f\"ur Chemische Physik fester Stoffe, 
D-01187, Dresden, Germany}
\author{H. Rosner}
\affiliation{Max Planck Institut f\"ur Chemische Physik fester Stoffe, 
D-01187, Dresden, Germany}
\author{S.-L. Drechsler}
\affiliation{Leibniz-Institut f\"ur Festk\"orper- und Werkstoffforschung
Dresden, D-01171, Dresden, Germany}

\date{\today}
\begin{abstract}
\noindent
We report on $^{6,7}$Li nuclear magnetic resonance measurements of the
spin-chain compound LiCu$_2$O$_2$ in the paramagnetic and magnetically
ordered states. Below $T$$\approx$24 K the NMR lineshape presents a
clear signature of incommensurate (IC) static modulation of the local
magnetic field consistent with an IC spiral modulation of the 
magnetic moments. $^{7}$Li NMR reveals strong phason-like dynamical
fluctuations extending well below 24 K.  We hypothesize that a series
of phase transitions at 24.2, 22.5, and 9 K 
reflects a "Devil's staircase" type behavior generic for IC systems.
LDA based calculations of exchange integrals reveal a large in-chain
frustration leading to a magnetical spiral.

\end{abstract}

 \maketitle

\noindent
Despite extensive 
efforts during the last decades spin ordering in frustrated
$S$=1/2 quantum spin chains still remains a matter of broad
activities.\cite{Zvyagin,choi,Hamada,krivnov,mizuno,matsuda}  
Rich phase diagrams with 
commensurate (C) and
incommensurate (IC) phases, with spin- and charge ordering, dimerization, or
superconductivity have been predicted.  
Most studies have been
focused on various cuprates
with corner- or edge-shared CuO$_4$ plaquettes.  Edge-sharing of
CuO$_4$ plaquettes leads to CuO$_2$ chains with a nearly 90$^\circ$
Cu-O-Cu bond angle causing a reduced nearest neighbor ($nn$) transfer
and  a next-nearest neighbor ($nnn$) transfer of similar size
allowing frustration effects. IC spiral states driven by ferromagnetic
(FM) $nn$ exchange and in-chain frustration 
have been predicted 
theoretically for CuO$_2$ chain compounds such as
Ca$_2$Y$_2$Cu$_5$O$_{10}$ but discarded 
experimentally.\cite{mizuno,matsuda}
The observation of 
in-chain IC effects in undoped quasi-1D
cuprates is so far restricted by a sharp {\it magnetic field driven}
C-IC transition observed in the spin-Peierls system CuGeO$_3$ at high
magnetic fields.\cite{Horvatic} Here we report
on $^{6,7}$Li nuclear magnetic resonance measurements (NMR) and local
density (LDA) based analysis of the electronic and magnetic structure
of the
chain compound LiCu$_2$O$_2$.  We show that the observed
spontaneous 
magnetic order can be described by a spiral modulation
of the
magnetic moments. Independently, LDA calculations and a
subsequent Heisenberg-analysis reveal strong 
in-chain frustration 
driving spiral ordering in accord with the NMR data.

\begin{figure}[h]
\includegraphics[width=8.0cm,angle=0]{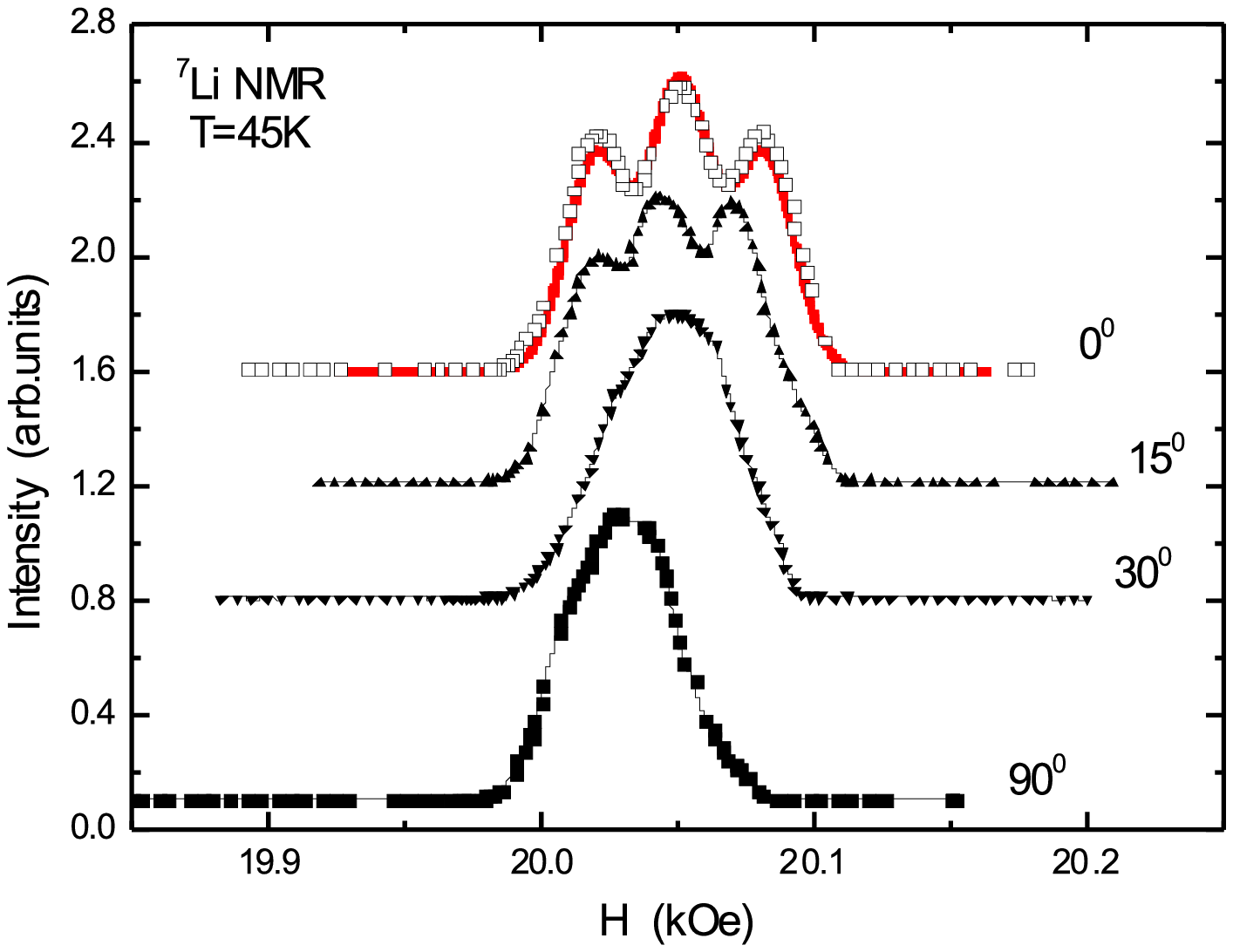}
\caption{The $^7$Li NMR spectrum  at $T$=45 K for different 
orientations of the magnetic field
with respect to the 
 {\bf c}-axis. The solid curve  on the 
upper spectrum: our simulation.} 
\label{fig2}
\end{figure}

LiCu$_2$O$_2$ is an insulating orthorhombic 
compound\cite{Hibble,Berger,Galakhov,Zvyagin}
with bilayers of edge-shared Cu$^{2+}$-O chains 
running parallel to the $b$-axis 
separated by Cu$^{1+}$ planes.
It exhibits
a high-temperature
antiferromagnetic(AFM)-like Curie-Weiss susceptibility $\chi
(T)$. Low-temperature $\chi (T)$ and specific heat 
stufies 
\cite{Roessli,Zvyagin} point to a series of intrinsic phase
transitions at $T$$\approx$24.2 K, $T$$\approx$22.5 K, and 
$T$$\approx$9 K pointing to a complex multi-stage rearrangement of the spin
structure. Magnetization studies
performed in external 
fields up to 5 T did not reveal any signatures of field-induced
transitions.  $\mu$SR 
data\cite{Roessli} point to a broad
distribution of magnetic fields at the muon stopping sites.  LSDA(+U)
calculations point to an FM
in-chain ordering.\cite{Galakhov} 
However, a simple FM ordering is in conflict with the
$\mu$SR data\cite{Roessli} and the AFM dimer liquid 
picture.\cite{Zvyagin,masuda}
Thus, to settle origin and character of the spin ordering
in LiCu$_2$O$_2$ further studies are required.
NQR and NMR measurements being a local probe 
are  standard tools to elucidate
the electronic and the magnetic structure.  
However, the
Cu NMR signal in LiCu$_2$O$_2$ stems presumably from
nonmagnetic Cu$^{1+}$ 
ions \cite{Fritschij} and reflects mainly
the disorder due to Li-Cu non-stoichiometry to be
discussed elsewhere.  We have performed $^{6,7}$Li NMR measurements on
large single crystals
\cite{crystal} at several temperatures both in the
paramagnetic\cite{Fritschij} and in the ordered phases.  
The spectra were measured for two main
orientations: ${\bf H}\parallel ({\bf a},{\bf b})$ and ${\bf H}\perp
({\bf a},{\bf b}$), and by 10$^\circ$ steps inbetween, by sweeping the
magnetic field at a fixed frequency of 33.2 MHz, the signal was
obtained by integrating the spin-echo envelope.
In the paramagnetic 
state ($T$=45K) the $^7$Li NMR line 
(Fig.\ \ref{fig2}) 
has a 
full width of 0.01 T (at the basement) and shows a
typical (for a $I$=3/2 nucle\textcolor{black}{us)} first 
order quadrupole splitting pattern with $\nu _{Q}$=51.7 kHz.
\begin{figure}[b]
\includegraphics[width=8.0cm,angle=0]{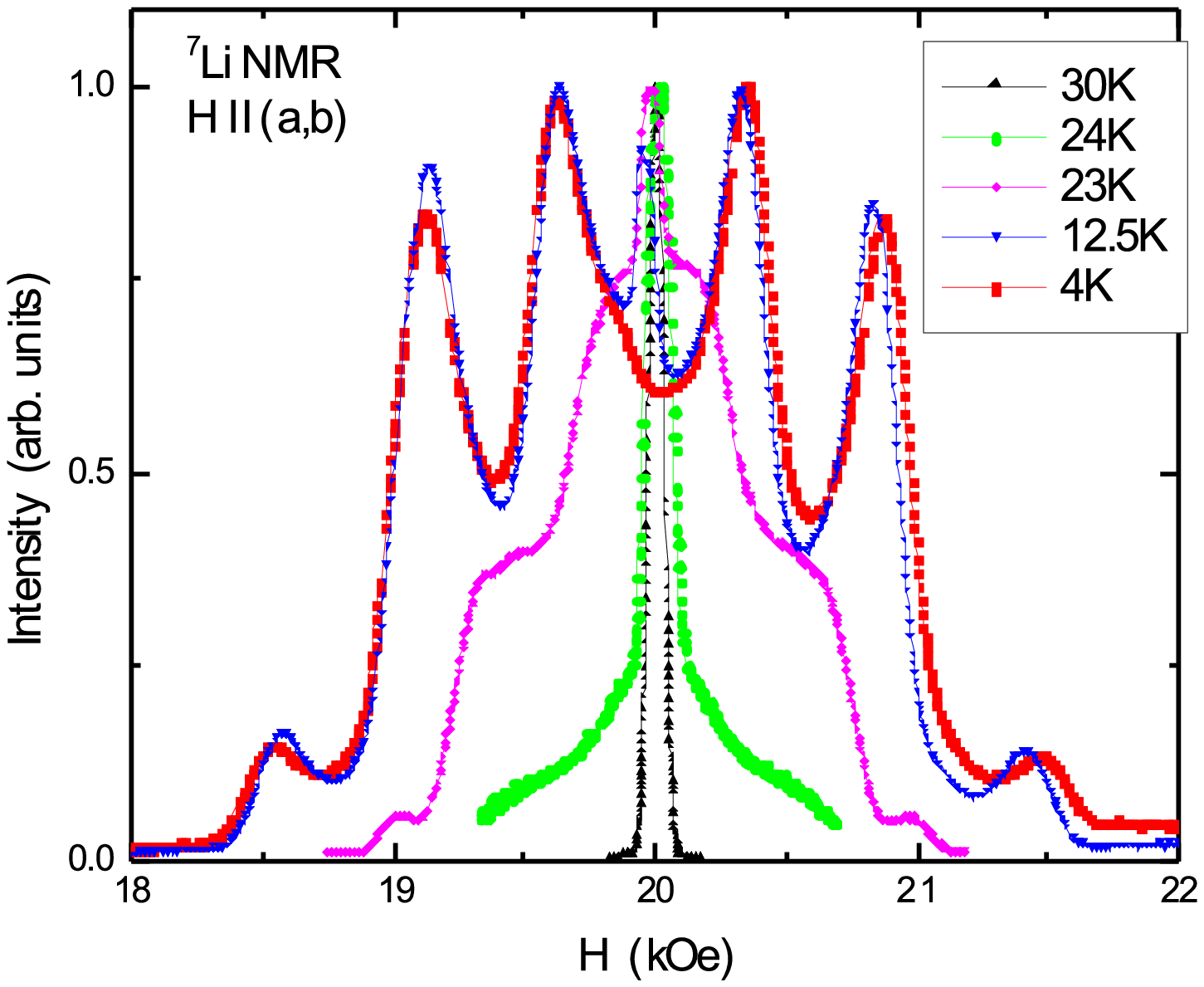}
\caption{The $^7$Li NMR spectra taken at various temperatures 
below the critical temperature of the magnetic ordering. The magnetic field 
is directed $\parallel$ 
to the ${\bf a}({\bf b})$ axis.} \label{fig3}
\end{figure}
 \textcolor{black}{ For $T$$<$24 K a dramatic 
change} of the 
$^7$Li NMR line shape is observed.  In general, the low-$T$  NMR 
line 
shape shows a 
wide continuous symmetric distribution which evolution is presented in 
Fig.\ \ref{fig3} for the ${\bf H}\parallel ({\bf a},{\bf b})$ geometry. 
In the ordered phase
we 
 still deal with a strong  and narrow 
central peak but with a well 
developed 
inhomogeneously broadened 
basement. Lowering $T$ the central peak weakens and 
disappears near $T$$\approx$10 K, while the basement gradually evolves  
into \textcolor{black}{a} well 
defined sextet \textcolor{black}{whose} full width exceeds by a 
factor \textcolor{black}{of} 30 (!) that of the paramagnetic phase. 
A similar situation takes place in   
${\bf 
H}\perp ({\bf a},{\bf b}$) geometry, but, 
with a specific quartet shape of the low-$T$
saturated NMR 
spectrum (Fig.\ \ref{fig4}).
Such  spectra are  
unique signatures of an infinite number of non-equivalent Li sites.
They correspond to an IC static modulation of local
fields
\cite{Berthier,Blinc, Blinc1} To
discriminate between magnetic and electric (quadrupole) effects we
have performed parallel $^6$Li NMR 
studies.\cite{remarkLi}  The
puzzling closeness of two NMR spectra below the critical temperature
(Fig.\ \ref{fig4}, top panel) unambiguously points to a distribution of
local magnetic fields as the origin of the observed
phenomena.
\begin{figure}[t] \includegraphics[width=8.cm,angle=0]{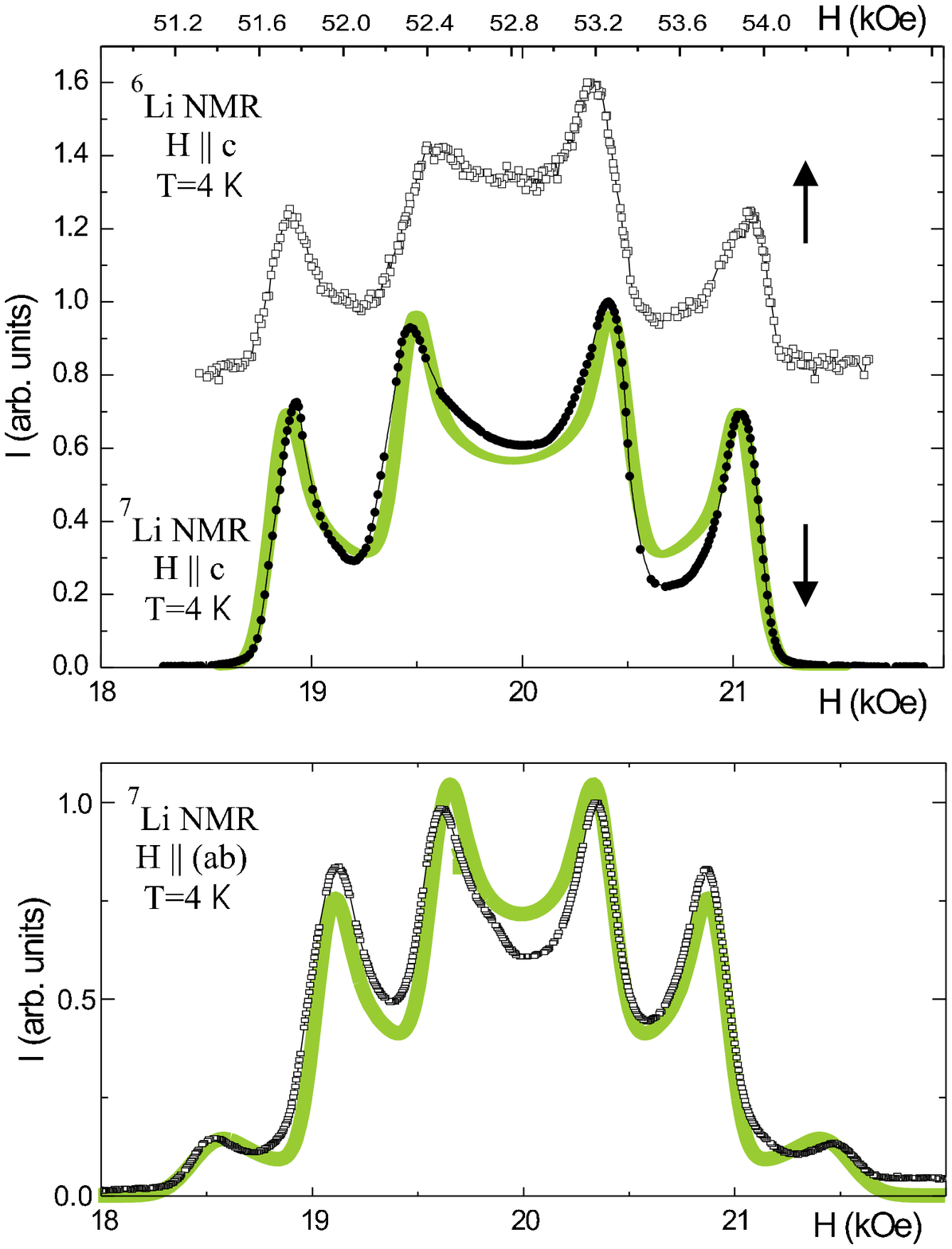}
\caption{The  $^6$Li  and $^7$Li  NMR 
spectra taken at 
4 K. 
 The solid curves in both panels show the results of our
theoretical simulation (see text).} 
\label{fig4}
\end{figure}
The  additional local magnetic field ${\bf h}({\bf R})$ seen
by a 
nucleus at a site ${\bf R}$ near a CuO$_2$  chain can be directly 
converted to a local spin polarization on the neighboring 
sites ${\bf S}({\bf 
R}+{\bf r})$: ${\bf h}({\bf 
R}) =\sum_{{\bf r}}{\hat A}({\bf r}){\bf S}({\bf R}+{\bf r})$, where ${\hat 
A}({\bf r})$ is the anisotropic tensor 
taking account of magnetic dipole and  
(super)transferred Li-Cu(O) hyperfine interactions.
When $h\ll H_0$ (the uniform external magnetic field: 
${\bf H}_0=H_0(\sin\Theta \cos\Phi , \sin\Theta \sin\Phi ,\cos\Theta $)):
$$
h=\frac{S}{2}
\sum_{{\bf r}} [A_{zz}\cos\Theta \cos\theta ({\bf R}+{\bf r})+
\sin\Theta \sin\theta ({\bf R}+{\bf r})
$$
\vspace{-0.7cm}
$$
[\cos\Phi \tilde A_{xx}\cos(\phi ({\bf 
R}+{\bf r})+\phi _1 )+
\sin\Phi \tilde A_{yy}\sin(\phi ({\bf R}+{\bf r})+\phi _2 )]],
$$
where $\tilde A_{xx}=[A_{xx}^2+A_{xy}^2]^{\frac{1}{2}}$, $\tilde 
A_{yy}=[A_{yy}^2+A_{yx}^2]^{\frac{1}{2}}$, $\phi _1 $ and $\phi _2 $ are phase 
shifts.
It should be noted that the IC structure can be described in different ways, 
both as a simple plane-wave \textcolor{black}{or} 
as a regular array of domain walls (solitons). 
If the concentration of solitons is large enough, we arrive at a soliton 
lattice.
Below we assume  a plane-wave modulated IC spin state, where the spin 
modulation phase is a harmonic function of the space coordinate in the 
direction of the modulation (hereafter Z-axis). Such a spiral spin structure we 
define as follows:
$
{\bf S}({\bf R})=S(\sin\theta ({\bf R})\cos\phi ({\bf R}), \sin\theta ({\bf 
R})\sin\phi ({\bf R}),\cos\theta ({\bf R})),
$
where  $\theta ({\bf R})=k_{1}Z + \theta _0 , \phi ({\bf R})=k_{2}Z + \phi _0$ 
with $k_{1,2}$ being IC spatial modulation wave numbers. In particular cases 
this 
spin structure can be reduced to any types of C states including 
simple FM- and AFM ones.
In the continuum approximation the resultant NMR line shape 
for a static case  
corresponds to the density of the local field distribution $g(Z)\propto 
|dH(Z)/dZ|^{-1}$ convoluted with the individual line shape.\cite{Blinc}
 Actually the straightforward application of this procedure is restricted by 
rather simple particular cases. 
However, the lack of a central peak and of any other sharp  lines in the NMR 
spectrum at $T$=4 K for all orientations of the external 
magnetic field points to a spatial modulation of all 
three spin 
components. 
Hence,
we are led to consider   complex spatial spirals with 
a 
modulation  of the longitudinal ($S_Z$) as well as of 
the transversal ($S_{X,Y}$) spin 
components. The specific quartet line shape  for the
${\bf H}\perp ({\bf a},{\bf b})$ geometry and the sextet one for ${\bf 
H}\parallel ({\bf a},{\bf b})$, with equivalent signals for 
${\bf H}_0 \parallel {\bf b}$, ${\bf H}_0 \parallel {\bf a}$, and the field oriented along the diagonal in $ab$-plane,
yield constraints for possible types of spirals     
reducing 
substantially their number. 
The measured NMR spectrum can be well described by
a simulation assuming a spin spiral twisting along 
the chain 
${\bf b}$ axis 
with identical modulation periods both for the 
longitudinal and the
transversal  spin components ($k_1$=$k_2$).
Within the 
long-wavelength  
limit and for a short-range 
hyperfine coupling 
 the NMR field  shift due to a (super)transferred hyperfine 
coupling in ${\bf H}\perp ({\bf a},{\bf b})$
 geometry for the spiral with a "positive" twisting along ${\bf b}$-axis 
(${\bf k}\parallel {\bf b}$, or $k>0$) can be written as follows:
$
h = (S/4) (\sum\tilde A_{xx})[\sin(2kZ+\phi _1 )-\sin\phi _1 ],
$
where $S\sum\tilde A_{xx}$ is the maximal value of the hyperfine field 
for the collinear spin ordering assumed. The spectral density function $g(H)$ has a rather simple form
$
g(H)\propto [1-(\omega (H)+\sin\phi _1 )^2 ]^{-\frac{1}{2}},
$
with $\omega (H)=4(H-H_0)/S\tilde A_{xx}$. As a result we arrive at two 
singularities $\omega _{\pm}(H)=\pm 1-\sin\phi _1 $ with a  separation $|H_+ 
-H_- |= (S/2) \tilde A_{xx}$ in between,  shifted with regard to $H_0$ 
by $(S/4) \tilde A_{xx}\sin\phi _1 $. Taking into account both 
"positively" ($k>0$) and "negatively" ($k<0$) twisted spirals we arrive at a
specific symmetric quartet NMR line shape. The situation looks slightly more 
complex for the "in-plane" geometry, where we deal with two types of twins. For
${\bf H}_0 \parallel {\bf b}$ or ${\bf a}$ axis we arrive at a
superposition 
of two different NMR signals stemming from differently oriented  twins, with 
${\bf H}_0 \parallel {\bf Z}$ or ${\bf H}_0 \perp {\bf Z}$, respectively.
The former are described by a simple expression for the NMR frequency shift:
$h = (S/2)A_{zz}\cos kZ $ with the well-known symmetric doublet
line shape \cite{Blinc}, while the latter yield a NMR response qualitatively 
similar to that of ${\bf H}\perp ({\bf a},{\bf b})$ geometry. 
Thus we 
arrive at a sextet line shape.  
 Making use of the Gauss convolution
procedure \cite{Blinc1} we successfully simulate the NMR spectra both for ${\bf H}\perp ({\bf a},{\bf b})$ 
and ${\bf H}\parallel ({\bf a},{\bf b})$ geometry (Fig.\ref{fig4}). It is 
noteworthy  that the current NMR study 
of IC phase does not yield information 
regarding both the IC period and the absolute value of the spin order 
parameter. Nevertheless, the simulation provides valuable
insight into 
the effective (super)transferred Cu(O)-$^{6,7}$Li hyperfine interactions.  
In fact, we obtain:
$S|\sum A_{zz}|\approx S|\sum \tilde A_{xx}|\approx$3 kOe,  
$S|\sum \tilde A_{yy}|\approx$2.6 kOe. 
The large magnitude and closeness  of these parameters point to the 
predominance of covalently enhanced isotropic (super)transferred Li-Cu(O) 
hyperfine interactions as compared with anisotropic magnetic dipole terms. 
It should be noted that our current approach implies  implicitly a  mutual 
collinear interchain ordering. A more detailed analysis of 
interchain coupling and impurity effects are 
challenging problem left for future studies.

The discreteness  of the crystal lattice causes 
pinning 
which tends to lock in the IC wave vector for a spiral at 
C values giving rise to a "Devil's staircase" type behavior. 
\cite{Blinc} 
We hypothesize that LiCu$_2$O$_2$ with a series of phase transitions at 24.2,
22.5, and 9 K exhibits such a behavior. This conjecture is supported by 
the temperature evolution of \textcolor{black}{our} 
NMR spectra (see Fig.\ref{fig3}) \textcolor{black}{which 
reveals a} well developed central peak in a wide temperature range below 24 
K. This is a signature of  strong low-energy dynamical fluctuations of local 
magnetic field typical for acoustic-like phase fluctuations, or phasons 
\textcolor{black}{generic for} IC structures. So, the  
  first order phase transition at 9 K
may be attributed to  pinning effects.
\textcolor{black}{Noteworthy the} phason-like spin fluctuations 
may result in a sizeable 
frequency dependence of the transverse relaxation rate T$_{2}^{-1}$ 
\textcolor{black}{affecting strongly}  the NMR line shape. These effects 
differ substantially for
different 
singularities in the NMR line shape, in particular, 
\textcolor{black}{they maybe} responsible for 
the 
weak intensity of the outer doublet in the 
NMR spectrum for ${\bf H}\parallel ({\bf 
a},{\bf b})$ geometry.
\begin{figure}[b]
\includegraphics[width=8.0cm,angle=0]{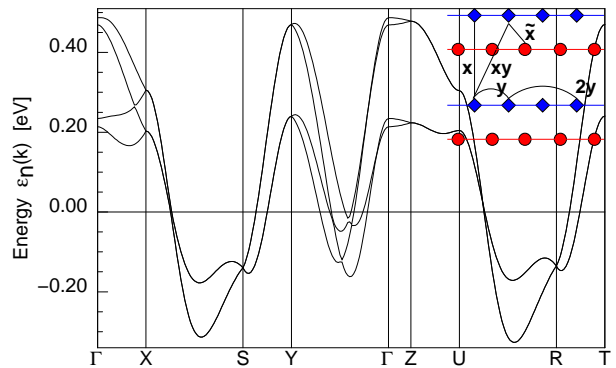}
\caption{ Band structure of LiCu$_2$O$_2$ near the Fermi energy, with
$\Gamma$=(0,0,0), X=(1,0,0), Y=(0,1,0), Z=(0,0,1), R=(1,1,1),
S=(1,1,0), T=(0,1,1), U=(1,0,1) in units of
$\pi/a$,$\pi/b$,$\pi/c$, respectively. The inset shows 
a sketch of a bilayer with double chains 
($\bullet$ and filled $\Diamond$) running along
{\bf b} together with our notation for the main
transfer and exchange processes.}
\label{band}
\end{figure}

To elucidate the origin of the observed IC spin ordering,
the knowledge of the relevant magnetic couplings is essential, 
especially
with respect to frustrations. In particular, the ratio
$\alpha$=$|J_{nnn}/J_{nn}|$ in a single-chain approach of 
the FM-AFM
$J_1$-$J_2$-Heisenberg model is decisive for C order ($\alpha<1/4$) or
IC one ($\alpha>1/4$).\cite{Hamada} To estimate
the main transfer integrals $t$ and exchange couplings $J$, we
performed full potential LDA 
calculations using the
FPLO code.\cite{FPLO} The resulting bands near the Fermi energy, 
shown in Fig.\ \ref{band}, 
have been fitted 
by an
extended tight-binding (TB) model to obtain the 
$t$'s.
The result is given in Tab.~\ref{tab}.  The in-chain dispersion along
the $k_y$ directions ($\Gamma$-Y) shows deep minima in the middle of
BZ indicating the predominance of the $nnn$ transfer over the $nn$ term
with $t_{2y}$$\approx$-109 meV and $|t_y|$$\approx$ 64 meV,
respectively.  This causes a strong in-chain frustration
 within the
spin-1/2 Heisenberg model 
$H=\sum_{<ij>}J_{ij}{\bf S}_i{\bf S}_j$.
The  AFM contributions to the $J$'s
estimated within a single band Hubbard model are given by
$J^{AFM}_{ij}=4t^2_{ij}/U$. 
%
We adopt for the screened on-site
repulsion $U$=3 eV as an appropriate value \cite{mizuno}. The result is given in
Tab.~\ref{tab}. 
\begin{table}[t]
\begin{tabular}{|l|c|c|c|c|c|}
\hline
&$y$&2$y$&$x$&$\tilde{x}$&$xy$\\ \hline
$|t|$&64&109&73&18&25\\
$J^{AFM}$&5.5&15.8&7.1&0.4&0.8\\
$J^{FM}$&-13.6&-1.4&-1.4&-&-\\
$J^{eff}$&-8.1&14.4&5.7&0.4&0.8\\ \hline
\end{tabular}
\caption{\label{tab} LDA derived transfer integrals $t$ and exchange
constants $J$ in meV, notation according to inset of Fig.~\ref{band}.}
\end{table}
 The screened
FM Heisenberg contributions
$J^{FM}$ (see Tab.~\ref{tab}) have been found 
using Wannier functions \cite{ku} of  Li$_2$CuO$_2$ which is
closely related to LiCu$_2$O$_2$ with respect to its 
in-chain geometry and
crystal field. The effective exchange $J^{eff}$ is given by the sum of
the two former contributions (see Tab.~\ref{tab}).
As one may expect for a nearly 90$^\circ$ Cu-O-Cu bond angle according
to the Goodenough-Kanamori rule, we arrive at a negative (FM)
total $nn$ in-chain exchange $J_{y}$$\approx$--8.1 meV,\cite{FM} and at an AFM
$nnn$ exchange: $J_{2y}$$\approx$ +14.5 meV.
The main interchain transfer process   $t_x$ 
is comparable
with the $nn$ term $t_y$, but the corresponding FM exchange contribution
is small
resulting in $J_x^{eff}$=5.7 meV. 
For the small transfer terms $t_{xy}$ and $t_{\tilde{x}}$, the
corresponding FM contributions are expected to be 
very small and are
ignored therefore.
From the values  in Tab.~\ref{tab} we  conclude that
LiCu$_2$O$_2$ is a quasi-2D system in an electronic sense but
magnetically more anisotropic.
%
Using the calculated $J$'s, we end up with $\alpha$$\sim$1.8 well
above the critical ratio of $1/4$. We emphasize that for a
3d arrangment of chains such as in LiCu$_2$O$_2$ and the
FM sign of $J_y$ found, the in-chain frustration is the {\it only} strong 
enough source to drive a spiral.
Thus LDA provides strong microscopic
support for the scenario suggested by the phenomenological NMR
analysis.  
%
In a more realistic anisotropic multiple chain situation, $\alpha_c$
would be slightly larger. However, with $\alpha$$\sim$1.8 the ignored
interchain coupling and spin anisotropy should not change  
essentially
our physical picture. This is at variance with the homogeneous 
FM in-chain order found in 
Ca$_2$Y$_2$Cu$_5$O$_{10}$ ($\alpha=0.05$) \cite{matsuda}  
and Li$_2$CuO$_2$ ($\alpha$$ <1$) with much stronger
interchain coupling. Hence, LiCu$_2$O$_2$ is unique, in being the first 
realization of a long sought for spiral chain compound. It is 
the most strongly
frustrated cuprate we know. For such a compound 
the study of doping and 
in particular the search of unconventional superconductivity  
are intriguing issues of general interest.

In conclusion, we present the first experimental evidence of a
low-temperature IC in-chain spin structure in an undoped quasi-1D
cuprate.\cite{remarklast} Below 24 K the $^7$Li NMR lineshape in
LiCu$_2$O$_2$ presents a clear signature of an IC static modulation
of the local magnetic field consistent with a spiral modulation of the
 magnetic moments twisted along the chain axis. The
intrinsic incommensurability accompanied by pinning effects is
believed to be responsible for an unconventional evolution of a
low-temperature spin structure with a series of successive phase
transitions resembling the "Devil's staircase" type behavior.
Applying an independent extended TB and subsequent Heisenberg analysis
based on LDA calculations we succeeded to elucidate a strong in-chain
frustration and this way also the microscopic origin of the magnetic
spiral seen by the NMR.
 
We thank A.\ Vasiliev, J.\ Richter, J.\ M\'alek, W.\ Ku, 
and H.\ Eschrig for fruitful discussions. We appreciate
support 
by grants MD-249.2003.02 [A.G.], SMWK, INTAS
01-0654, CRDF REC-005, E 02-3.4-392, UR.01.01.042), RFBR 01-02-96404
[A.M.], DFG SPP 1073, Es-85/8-10 [S.D.], and DFG
Emmy-Noether-program [H.R.]


\begin{thebibliography}{99}
\bibitem{FM}
A similar conclusion has been drawn in Ref.\onlinecite{Galakhov}
\bibitem{remarklast}
After the submission of our manuscript, we learned about 
neutron 
data \cite{masuda} also pointing to 
a spiral  
propagating  along $b$. 
However, their
interpretation in terms of a dimer liquid \cite{Zvyagin,choi}
is based on pure 
AFM $J$'s and a planar ($ab$) spin structure 
is inconsistent with our results.
\end{thebibliography}
\end{document}